# Protein–DNA computation by stochastic assembly cascade

Roy Bar-Ziv, Tsvi Tlusty, and Albert Libchaber*

Center for Physics and Biology, The Rockefeller University, 1230 York Avenue, New York, NY 10021

**The assembly of RecA on single-stranded DNA is measured and interpreted as a stochastic finite-state machine that is able to discriminate fine differences between sequences, a basic computational operation. RecA filaments efficiently scan DNA sequence through a cascade of random nucleation and disassembly events that is mechanistically similar to the dynamic instability of microtubules. This iterative cascade is a multistage kinetic proofreading process that amplifies minute differences, even a single base change. Our measurements suggest that this stochastic Turing-like machine can compute certain integral transforms.**

Every computation requires a reliable recognition of its input data. Any scheme for computation based on protein–DNA binding must attain this recognition within the physical properties of this interaction, its specificity, affinity, and cooperativity. These properties define biochemical networks such as those used by the cell to process information received from stimuli and to compute its response. The resulting computations are inherently stochastic due to the "noisy" nature of biochemical pathways that resemble more a probabilistic pinball machine than a deterministic desktop PC.

So far, artificial, *in vitro* biomolecular computing strategies relied mainly on Watson–Crick complementarity of DNA (or RNA). These schemes, which were used to solve a certain class of hard-to-compute problems, are almost deterministic due to the relatively high hybridization energy. The typical algorithm, combinatorial search, encodes potential solutions as DNA sequence library and then selects correct solution(s) via parallel filtration, eliminating the wrong solutions by manipulations based on complementarity (1, 2). Another approach constructs finite-state computing machines, the internal states of which are encoded in DNA sequence (3).

Here we report an *in vitro* stochastic biomolecular computation based on low-specificity protein–DNA binding: An assembly cascade of RecA proteins on single-stranded DNA can discriminate between similar sequences, thus fulfilling a basic computational task that may be one stage in a more complex computation. The assembly process overcomes the error-prone nature of the single protein binding by constructing a multistage cascade, similar to kinetic proofreading (4), in which many proteins bind and unbind collectively. We find that the dynamics of the cascade is mechanistically similar to the dynamic instability of microtubules, which is used as an efficient space search algorithm within the living cell (5). It also resembles a stochastic counter (6), an imperfect digital apparatus that registers the number of certain events (think of a voting machine). The collective, nonlinear mode of operation of the cascade enables sensitive discrimination of minute length and sequence differences including a single base change.

The hardware of our molecular machine comprises a test-tube filled with a solution of single-stranded DNA molecules, RecA proteins, and ATP molecules that fuel the assembly cascade. When the concentration of RecA monomers exceeds some onset value, they start to form helical filaments, one RecA monomer per each base triplet, that envelope and stretch the DNA (7). A filament first forms when a nucleus, a RecA monomer, binds to a random site along the DNA and then extends rapidly by polymerization to the 3′ end of the empty strand. When bound to DNA, RecAs hydrolyze ATP and change their conformation into a less stable state. The RecA that is closest to the 5′ end, with only one neighboring monomer, tends to disassemble back into the solution when hydrolyzing ATP (8). The resulting assembly–disassembly cascade is asymmetric; while nucleation events extend the filament by long chunks, disassembly removes monomers one by one. A graphical manifestation of this stochastic asymmetry is the irregular saw-tooth form of the filament length (or machine state) dependence on time (Fig. 1A).

Rather than further describing the extensively studied biochemistry of RecA assembly (7) we focus on the computational features of this protein–DNA molecular machine, its "software." We use here the notion of "machine" in the sense of certain physical realization of an abstract computation, sequence discrimination in our case. Nucleation and disassembly are the two basic operations of this machine. They change the machine's internal state, which is determined by the current length of RecA filament. To describe the machine dynamics, we use the traditional state-transition diagram, where circles represent states, and arrows represent transitions between states (Fig. 1B). In state $Q_n$, $n$ binding sites out of total $N$ sites along the DNA are vacant, and the RecA filament length is therefore $N - n$ ($Q_0$ is a fully covered DNA, and $Q_N$ is an empty strand). Clearly, this is a *finite-state machine* with the number of states equal to the number of binding sites, $N$. The symbols on each arrow represent the probability per unit of time that such transition occurs given that the machine is in the state at the tail of the arrow. Disassembly can take the machine from state $Q_n$ to the next state of the cascade $Q_{n+1}$ at rate $\kappa_-$ whereas at nucleation events the machine jumps from $Q_n$ to any of the lower states $Q_m$, $m < n$, at rate $\kappa_+$. We also need an output device that will report the machine's current state. In the experiment, the molecular machine "reports" its state through a change in the rotational motion of the DNA molecule, which is directly related to the number of bound RecA monomers and measured by fluorescence anisotropy (9).

The stochastic state-transition diagram can be expressed as a set of $N$ differential equations for the probabilities $p_n$ that the machine is at state $Q_n$. Summing the incoming (first two terms) and outgoing (last two terms) transitions at each state of the diagram we obtain

$$\frac{dp_n}{dt} = \kappa_- p_{n-1} + \kappa_+ \sum_{m=n+1}^{N} p_m - \kappa_- p_n - \kappa_+ n p_n. \qquad [1]$$

The state-transition diagram couples each polymerization state $Q_n$ with all the lower states $Q_m$, $m < n$ (Fig. 1), and the equivalent master equation (Eq. **1**) is therefore integro-differential, with boundary and normalization conditions $dp_N/dt = \kappa_- p_{N-1} - Np_N$, $\Sigma_{n=0}^{N} p_n = 1$. To reduce the connectivity of the state-transition diagram, we express it in terms of the cumulative $P_n = \Sigma_{m=n}^{N} p_m$, the probability to find the machine at a stage higher or equal to $Q_n$. It is also the probability that the filament is shorter than $N - n$ and that site $n$ is empty. Two processes can alter $P_n$: (*i*) disassembly fronts vacate site $n$ at a rate proportional to the front position distribution, $-p_n = P_n - P_{n+1}$, and (*ii*) nucleation at any of the $n$ vacant sites fills site $n$. The dynamics is simpler, since any possible nucleation from a higher state, $Q_m, m > n$ leaves $P_n$ unchanged. The resulting master equation is local, $dP_n/dt = -\kappa_-(P_{n+1} - P_n) - \kappa_+ n P_n$, with the normalization $P_0 = 1$. Technically, one can obtain this result directly by summing Eq. **1** from $n$ to $N$ with the boundary conditions.

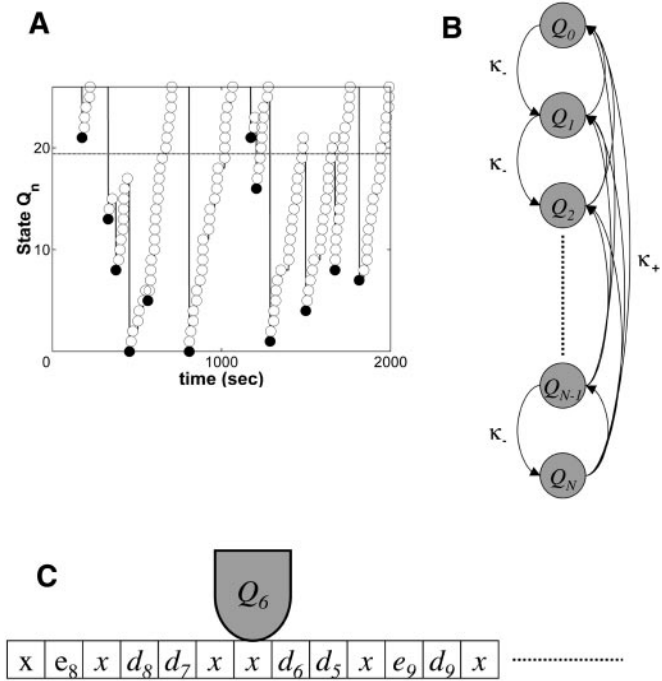

**Fig. 1.** (*A*) Simulation of a 26-state stochastic assembly cascade. State $Q_n$ is a RecA filament of length $26n$. The assembly machine advances to higher states through protein disassembly at the filament end (open circles) and to lower states by nucleation (solid circles). The machine's state fluctuates strongly around the ensemble average (dashed). (*B*) State-transition representation of the machine dynamics. The interplay between $N$ disassembly steps at rate $\kappa_-$ (*Left*) and $N(N+1)/2$ nucleation paths at rate $\kappa_+$ (*Right*). (*C*) A deterministic reading head changes its internal state according to the tape it reads square by square from left to right. The tape is produced in a random process that maps the stochasticity of the assembly (see text).

Thinking of $n$ as a spatial coordinate, we approximate the discreet master equation for $P_n(t)$ by a "drift" equation for the continuous cumulative probability $P(n, t)$.

$$\frac{\partial P(n)}{\partial t} = -\kappa_- \frac{\partial P(n)}{\partial n} - \kappa_+ n P(n) \quad [2]$$

The disassembly term is approximated by a gradient, neglecting higher order derivatives in the Kramers–Moyal expansion (10). In particular, we omit the familiar second-order diffusive term that plays a minor role as long as disassembly is much faster than nucleation rate, in the regime, $(\kappa_+/\kappa_-)N \ll 1$, which includes the large fluctuation regime, $(\kappa_+/\kappa_-)N^2 \approx 1$, where the RecA-assembly cascade is the most sensitive.[†]

---

[†]We note that the assembly dynamics differs essentially from the Langevin dynamics of a particle diffusing in a one-dimensional random force field (11) or the related asymmetric exclusion process (12): While a diffusing particle travels continuously, the filament end can abruptly jump to a new site by nucleation. The master equation, therefore, does not lead to the familiar Fokker–Planck equation, and an equivalent Langevin formulation would require infinite stochastic forces to enable the nucleation jumps (10). The effect of

Master equations such as **1** and **2** are generic in stochastic transition processes, especially in chemical kinetics (10). What makes the computing-machines terminology natural in our case is the understanding that the *RecA-binding cascade processes information encoded in the DNA sequence*. This may be clarified if one considers a concrete machine model of the cascade. This time we think of a Turing-like device, a deterministic machine that is coupled to an infinite tape through a reading head (Fig. 1*C*). The internal states of the machine are the same $N$ binding states $Q_n$. The noisy Brownian dynamics of the cascade is embedded in the tape, which is produced by the following procedure: Time is divided into an infinite series of short equal segments that correspond to the squares on the tape. To each square we randomly assign a symbol with a probability that matches the transition rates. We denote disassembly from state $Q_n$ by $d_n$, nucleation to state $Q_n$ by $e_n$, and in the rest of the squares we write $x$ to denote that nothing happens during the corresponding time duration. The machine reads the squares sequentially from, say, left to right and responds according to the symbol written in the current square. Suppose that the machine is at state $Q_n$, then it responds according to a simple set of rules. (*i*) If it reads $d_n$, it moves to state $Q_{n+1}$. (*ii*) If it reads $e_m$ and $m < n$ the machine moves to state $Q_m$. (*iii*) In all other cases, if it reads $x$ or $d_m$ with $m \neq n$, or $e_m$ with $m \neq n$, then it stays at state $Q_n$. After its state is determined, the reading head moves one square to the right.

Stochastic automata are natural to information processes ever since they emerged in Shannon's classical study of communication channels (14, 15). The notion of stochastic computers was introduced to the molecular realm in Bennett's discussion of DNA translation and replication, where the computational task is sequence copying (16). We show below that rather than Xeroxing the sequence like RNA- and DNA-polymerase, the RecA cascade carries out another type of computation, the discrimination of close-by sequences. Sequence information is encoded in the random tape through the dependence of the probabilities for disassembly ($d_n$) and nucleation ($e_n$) events at a certain site $n$ on the specific base triplet. This information can be equivalently encoded as sequence-specific transition rates, $\kappa_-(n)$ and $\kappa_+(n)$, in the state-transition diagram and the corresponding master equation. Although RecA is a nonspecific binding protein with similar affinities for many possible triplets, our measurements show (9) that the collective assembly cascade constructed from these low-specificity components is a highly specific detector that can amplify and discriminate even minute sequence differences (17).

The "sequence-detector" machines we construct are assembly cascades on single-stranded DNAs, 39 or 78 bases long (13- and 26-stage machines). Any measurement that tries to "look inside" such a stochastic machine, that is to infer its internal dynamics from observable output, has to rely on statistical analysis (18). One must collect a sufficient set of observations to overcome the noisiness of the output. We resolve this difficulty by simultaneously measuring many identical machines, $\approx 10^5$–$10^6$ fluctuating DNA–RecA complexes that produce a very smooth ensemble-average signal. An alternative approach could be time averaging over a single-molecule signal (19). Our DNAs carry a fluorescent dye attached to their 3′ end. The fluorescence anisotropy of the dye reports RecA binding as it slows down the rotational motion of the DNA (9). The response of the cascade is examined as we tune the interplay between nucleation and disassembly by changing the available amount of RecA in the solution. The nucleation rate

---

diffusion remains minor for short enough inhomogeneous sequences such as the sequence with the point mutation measured in the experiment. In contrast, assembly on longer sequences, $(\kappa_+/\kappa_-)N \approx 1$, is predicted to exhibit a striking randomness effect of both sequence and diffusion, which may lead to anomalous motion and phase transitions (13).

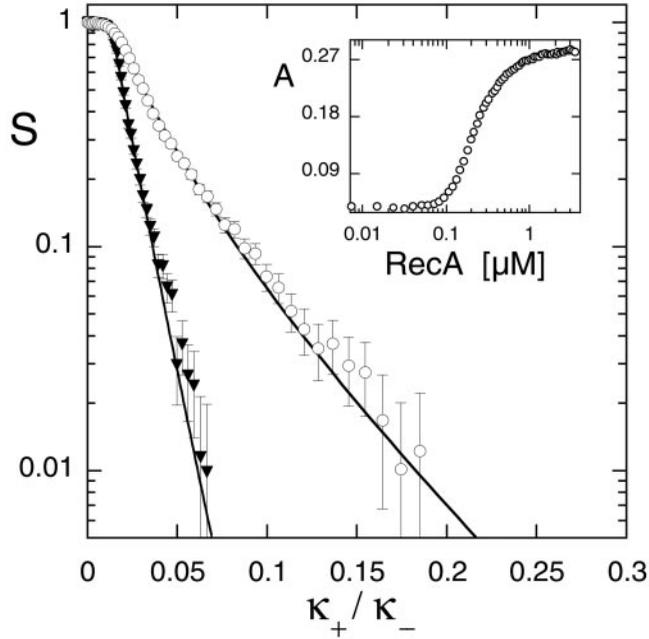

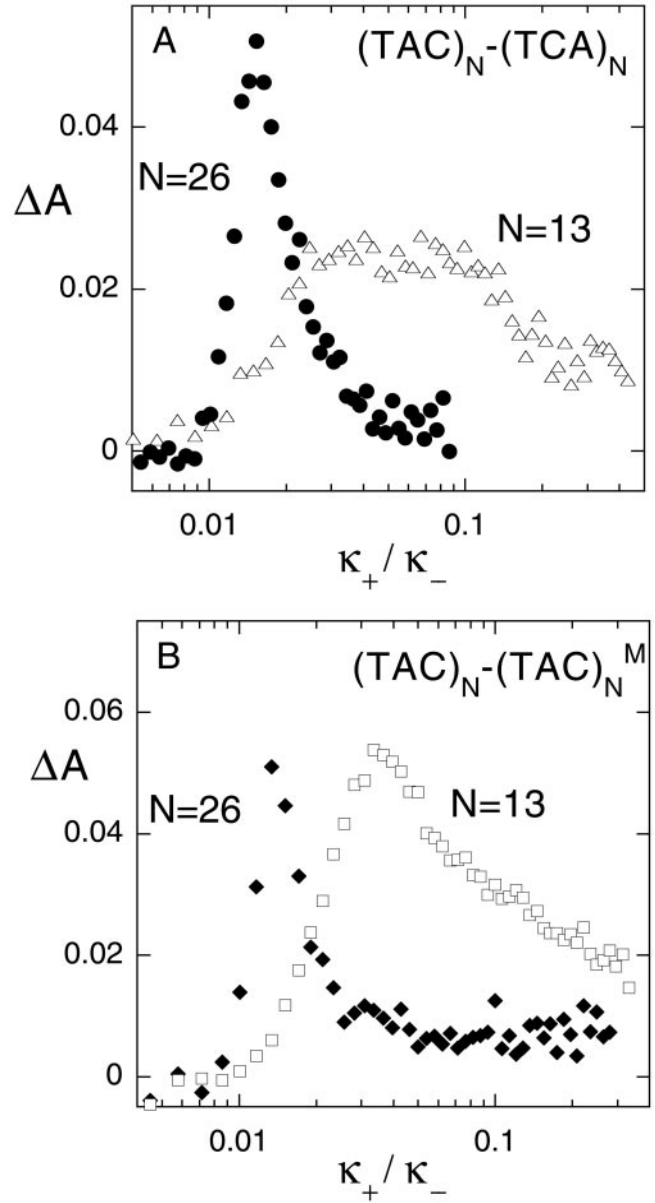

**Fig. 2.** Increase of fluorescence anisotropy, $A$, upon RecA binding on $(TAC)_{13}$ measured in steady state as a function of RecA concentration $R$ (*Inset*). Thermal rotation of naked DNAs decreases the intrinsic anisotropy of the dye, $A_m \sim 0.25 - 0.31$, to $A_0 \sim 0.05$, as reported previously (9). By varying sequence length, we found that $A$ increases linearly with RecA binding near the 3' end before saturating at $A_m$ with $l_0 = 7$ RecA monomers (unpublished data). The anisotropy therefore measures an average over the cascade occupancy, $A = A_m - (A_m - A_0)S; S = 1/l_0 \int_{N-l_0}^{N} P(n)dn$. Independent kinetic measurement indicated $\kappa_T = \kappa_+/R = 6.4 \times 10^{-3}$ sec$^{-1}\cdot\mu$M$^{-1}$ and $\kappa_- = 4.4 \times 10^{-2}$ sec$^{-1}$. The normalized anisotropy, $S$, is plotted as a function of the rate ratio $\kappa_+/\kappa_-$. Both curves, $(TAC)_{13}$ (open circles) and $(TAC)_{26}$ (triangles), decay exponentially close to saturation with a higher slope for the longer sequence that has more cascade stages (solid line). DNA sequences were synthesized, labeled with 3'-fluorescein (Midland Certified Reagent, Midland, TX), and HPLC-purified. The DNA concentration, 39 nM in nucleotide (13 nM RecA-binding sites), was much lower than the RecA binding onset (70 nM) and ensured that the free RecA was in practice the total RecA. RecA protein (New England Biolabs) binding assays were done in 25 mM Tris·HCl, pH 7.5/150 mM NaCl/1 mM MgCl2/1 mM DTT/1 mM ATP. Fluorescence was excited in a quartz cuvette (3 ml) by a vertically polarized 488-nm argon laser line. Emission intensity polarized parallel, $I_v$, and perpendicular, $I_p$, to the excitation was measured (9), from which the anisotropy was determined: $A = (I_v - I_p)/(I_v + 2I_p)$.

**Fig. 3.** (*A*) The difference between the fluorescence anisotropy signal of $(TAC)_N$ and $(TCA)_N$ for $n = 13,26$. Sequence separation peaks at a lower rate ratio, $\kappa_+/\kappa_-$, for longer sequences in agreement with cascade model. (*B*) A single base change was introduced at the seventh triplet of $(TAC)_N$: $(TAC)_{13} \rightarrow (TAC)_{13}^M = (TAC)_6(TAG)(TAC)_6$, $(TAC)_{26} \rightarrow (TAC)_{26}^M = (TAC)_{19}(TAG)(TAC)_6$. This change induces a weak hairpin secondary structure in the DNA, which forms a barrier for RecA binding (9) and shifts the binding curves of $(TAC)_N^M$ toward higher RecA concentration with respect to $(TAC)_N$. The difference in signal between $(TAC)_N$ and $(TAC)_N^M$ is peaked at a lower rate ratio for longer sequences.

at any vacant site increases with RecA concentration, $R$, like $\kappa_+(n) = \kappa_T(n)\cdot R$ (where $\kappa_T(n)$ is the triplet-specific rate constant), while the disassembly rate remains constant as the amount of ATP it consumes is kept at saturation level. When we add RecA to the test tube more monomers bind DNA through nucleation-polymerization process, and the chance to find occupied sites increases. The binding curve is sigmoidal, typical of *collective* chemical kinetics (Fig. 2 *Inset*).

Our measurement suggests an exponential sensitivity of the assembly cascade to sequence length and RecA-binding rate constants, $\kappa_-$ and $\kappa_+$ (Fig. 2). To understand how such amplified sensitivity is accomplished, we reexamine the master equation (Eq. **2**). Motivated by our measurements that indicate fast relaxation of the assembly cascade, we study its steady state. For a uniform sequence, $\kappa_-(n) = \kappa_-$, $\kappa_+(n) = \kappa_+$, the steady-state probability distribution is Gaussian,

$$P(n) = \exp\left(-\frac{\kappa_+}{2\kappa_-}n^2\right). \quad [3]$$

It follows that even a slight difference in transition rates of two uniform sequences is exponentially amplified as the cascade advances to its higher states (large $n$). The maximal enhancement increases exponentially like the square of the states number, actually the number of DNA base triplets. Similarly, the cascade can discriminate between lengths, $N_1$ and $N_2$, of two sequences made of the same triplets, since the probability ratio at the highest states is $P(N_1)/P(N_2) = \exp[-(\kappa_+/2\kappa_-)(N_1^2 - N_2^2)]$. The exponential amplification is the result of the *iterative, multistage* structure of the cascade. It is the same design principle that underlies industrial distillation (20) and the kinetic proof-reading pathway of protein synthesis (4). The exponential amplification of the cascade is evident from the behavior of the normalized fluorescence anisotropy of uniform triplet repeats at the saturated regime (Fig. 2). In this regime of high nucleation rate, it becomes harder for the machine to climb up to higher states through successive disassembly steps (Fig. 1*B*). However, this helps the cascade to discriminate lengths, because shorter sequences need less disassembly steps to reach higher states, and indeed the curve for the longer (TAC)$_{26}$ triplet-repeat sequence is steeper than that of the half-size (TAC)$_{13}$.

We test the sequence-discrimination capability of the cascade by comparing the binding curves of two uniform single-stranded DNA molecules made of very similar triplet repeats, TAC and TCA (Fig. 3*A*). The difference between the two binding signals behaves similarly to the relative entropy of the two probability distributions (sometimes called "information for discrimination"; ref. 21) and therefore gives a good idea about their distinguishability. For both lengths $n = 13,26$ we find that the difference peaks at a certain rate ratio $\kappa_+/\kappa_-$ that corresponds to the maximal slope of the binding curves (Fig. 2 *Inset*), where cooperativity is highest (Eq. **2**). The peak indicates optimal tuning of the back and forth scanning motion that is used by the stochastic machine to "read" the sequence (a process that was mapped to sequential reading of a random tape). Thinking of the serriform time series (Fig. 1*A*) as a "sentence" composed of an *N*-state alphabet printed by a stochastic typewriter (something like . . . $Q_{10}Q_{11}Q_8Q_9Q_3Q_3Q_3Q_4$. . . ), then the appearance of the "letter" $Q_N$ corresponds to a completed scan. Interpreting $Q_N$ as a "space bar," the maximal rate of completed scans corresponds to the most informative reading with the maximal rate of "words." This occurs at the "working point" of the cascade, $t_+/t_- \sim 1$, when the time interval between nucleation events, $t_+ \sim 1/(N\kappa_+)$, is matched with the time required to climb back to state $Q_N$, $t_- \sim N/2\kappa_-$. With the rates measured independently by kinetic assays we find that optimal separation occurs indeed in the optimal regime, $t_+/t_- \sim 1$–3.

The protein assembly cascade dynamics can detect also localized differences in nonuniform sequences. A stringent test for our machine is the discrimination of a single base change. We therefore introduced a change C → G at the seventh triplet of the two uniform (TAC)$_N$ sequences and measured the discrimination (Fig. 3*B*). Similar to the uniform sequences, the difference in binding between a sequence and its variant peaks at an optimal $\kappa_+/\kappa_-$ that is lower for the longer sequence, consistent with the working point.

The machine's ability to discriminate localized changes suggests a basis for certain mathematical computations, integral transforms. Consider an ensemble of uniform sequences made of *N* RecA-favored triplets (relatively high $\kappa_+$; ref. 9). Within each sequence, we encode a "defect" in the form of a single unfavorable triplet placed at one of the *N* possible sites. Let $w(n)$ designate the fraction of sequences with defect at site *n*. A test tube with a mixture of all these sequences encodes the vector $[w(1), w(2), \ldots w(N)]$. A monodisperse solution with defect only at site *n* is one of the *N*-unit base vectors that span our sequence space. As shown below, the signal from such a base vector is exponential in the position of the defect, $S(n, k) \sim \exp(-k n)$, where $k = \Delta(\kappa_+/\kappa_-)$ is the difference in the ratio of reaction rates at the defect. Since the fluorescence anisotropy is an ensemble average, the signal of mixture is a Laplace-like transform,

$$S(k) = \sum_{n=0}^{N} w(n)S(n,k) \sim \sum_{n=0}^{N} w(n)\exp(-kn). \quad [4]$$

To account for the nonuniformity of a DNA sequence with site-dependent nucleation and disassembly rates, $\kappa_+(n)$ and $\kappa_-(n)$, we modify the continuous master equation to

$$\frac{\partial P(n,t)}{\partial t} = -\frac{\partial}{\partial n}[\kappa_-(n)P(n,t)] - \left(\int_0^n \kappa_+(m)dm\right)P(n,t),$$

with the inhomogeneous steady-state solution

$$P(n) = \frac{\kappa_-(0)}{\kappa_-(n)} \exp\left(-\int_0^n \frac{dm}{\kappa_-(m)}\int_0^m \kappa_+(q)dq\right).$$

A mutation at site $n_0$ implies a localized change of reaction rates by $\Delta\kappa_+$ and $\Delta\kappa_-$. When the mutation is in a formerly uniform sequence, variation of the steady-state profile exhibits a change that depends on the position of the site as $\Delta P(n)/P(n) \simeq -\kappa n_0$, where the "wave number," $k$, is the difference in the reaction rates ratio, $k = \Delta(\kappa_+/\kappa_-) = (\kappa_+/\kappa_-)(\Delta\kappa_+/\kappa_+ - \Delta\kappa_-/\kappa_-)$. The resulting relative change in $P(n)$ depends exponentially on the position of the mutation $n_0$. Integrating over $P(n)$ we find that the anisotropy signal scales like $S(n_0, k) \sim \exp(-kn_0)$.

By choice of other types of sequence base vectors, the stochastic cascade machinery, through the ensemble measurement, can encode and decode mixtures in terms of other transforms. It is tempting to speculate that with additional operations to manipulate sequences at hand, such as recombination, one could construct a molecular architecture for more complex computations. The question of whether RecA assembly is used for natural computation requires *in vivo* testing (22).

We thank K. Adzuma and B. Shraiman for discussions and suggestions and D. Thaler for a fruitful and inspiring collaboration.

1. Adleman, L. M. (1994) *Science* **266,** 1021–1024.
2. Landweber, L. F. & Kari, L. (1999) *BioSystems* **52,** 3–13.
3. Winfree, E., Liu, F. Wenzler, L. A & Seeman, N. C. (1998) *Nature (London)* **394,** 539–544.
4. Hopfield, J. J. (1974) *Proc. Natl. Acad. Sci. USA* **71,** 4135–4139.
5. Mitchison, T. & Kirschner, M. (1984) *Nature (London)* **312,** 232–242.
6. Killeen, P. R. & Taylor, T. J. (2000) *Psychol. Rev.* **107,** 430–459.
7. Kowalczykowski, S. C., Dixon, D. A., Eggelston, A. K., Lauder, S. D. & Rehrauer, W. M. (1994) *Microbiol. Rev.* **58,** 401–465.
8. Shan, Q., Bork, J. M., Webb, B. L., Inman, R. B. & Cox, M. M. (1997) *J. Mol. Biol.* **265,** 519–540.
9. Bar-Ziv, R. & Libchaber, A. (2001) *Proc. Natl. Acad. Sci. USA* **98,** 9068–9073.
10. van Kampen, N. G. (2002) *Stochastic Processes in Physics and Chemistry* (Elsevier, Amsterdam).
11. Bouchaud, J. P, Comtet, A., Georges, A. & Le Doussal, P. (1990) *Ann. Phys.* **201,** 285–341.
12. Schutz, G. M. (1997) *J. Stat. Phys.* **88,** 427–452.
13. Lubensky, D. K. & Nelson, D. R. (2000) *Phys. Rev. Lett.* **85,** 1572–1575.
14. Shannon, C. E. (1948) *Bell Syst. Tech. J.* **27,** 379–423.
15. Shannon, C. E. (1948) *Bell Syst. Tech. J.* **27,** 623–656.
16. Bennett, C. H. (1982) *Int. J. Theor. Phys.* **21,** 905–940.
17. von Neumann, J. (1956) in *Automata Studies*, eds. Shannon, C. E. & McCarthy, J. (Princeton Univ. Press, Princeton), pp. 43–98.
18. Grenander, U. (1966) *Res. Pap. Statist. Testchrift J. Neyman* 107–123.
19. Hegner, M., Smith, S. B. & Bustamante, C. (1999) *Proc. Natl. Acad. Sci. USA* **96,** 10109–10114.
20. Lord Rayleigh (1896) *Philos. Mag.* **42,** 493–498.
21. Cover, T. M. & Thomas, J. A. (1991) *Elements of Information Theory* (Wiley, New York).
22. Matic, I., Rayssiguier, C. & Radman, M. (1995) *Cell* **80,** 507–515.